\documentclass{aa}
\usepackage{graphicx}

\begin{document}

   \title{High NH$_2$D/NH$_3$ ratios in protostellar cores}

   \subtitle{}

   \author{J. Hatchell
          \inst{1}
          }

   \offprints{J. Hatchell}

   \institute{Max-Planck-Institut f\"ur Radioastronomie, Auf dem H\"ugel 69, 53121 Bonn, Germany\\
              \email{hatchell@mpifr-bonn.mpg.de}
             }

   \date{}

\abstract{ Observations of low mass protostars which probe small
enough size scales to be within likely CO depletion regions show the
highest [NH$_2$D]/[NH$_3$] ratios yet measured, of 4--33\%.  These
molecular D/H ratios are higher than those measured on larger scales,
showing that deuterium fractionation increases towards
protostellar cores.  As in cold clouds, such high ratios can be
produced by gas-phase ion-molecule chemistry in the presence of
depletion.  Grain surface chemistry is less likely to explain the
deuterium enhancement, 
as it would require 
higher fractionation in ices than current
models 
predict.  The link between
accretion, depletion and high molecular deuterium fractionation is
strongly supported.}  \maketitle


\section{Introduction}
%
There are two main ways of producing molecular D/H enhancements in
molecular clouds over the $\sim 10^{-5}$ ratio of $[{\rm HD}]/[{\rm
H}_2]$.  Firstly, grain surface chemistry may enhance molecular D/H
ratios (Brown \& Millar~\cite{brownmillar89a},\cite{brownmillar89b};
Charnley et al.~\cite{charnley97}).  Secondly, some key gas phase
reactions involving destruction of deuterated species run slower at
low temperatures than the equivalent reactions with hydrogen, and this
leads to molecular D/H enhancements where a cold gas phase chemistry
has been active.  Furthermore, in colder gas, depletion of heavy
molecules such as CO results in an increase of
[H$_2$D$^+$]/[H$_3^+$] and molecular D/H ratios (Brown \&
Millar~\cite{brownmillar89a}; Roberts \&
Millar~\cite{rmsingled},\cite{rmdoubled}; Rodgers \&
Charnley~\cite{rcnd2h}).

Depletion of heavy molecules during freezeout in cold molecular clouds
can result in molecular D/H ratios above 10\% (Roberts \&
Millar~2000), and can explain many of the high ratios observed in
cold, dark clouds: DCO$^+$ fractionation in L1544 and L134N (Caselli
et al.~\cite{caselli99}; Tin\'e et al.~\cite{tine00}) and ND$_2$H in
L134N (Roueff et al.~\cite{roueff00}).  

Young protostars should show the effect of freezeout as well as cold
dark clouds, though grain mantle release may also influence the
ratios, as in IRAS~16293$-$2422 (van Dishoeck et
al.~\cite{vandishoeck95}).  A few studies have looked at molecular D/H
ratios in 
low mass protostars: HDCO and DCN (Roberts
et al.~\cite{roberts02}); NH$_2$D (Shah \& Wootten~\cite{sw01}; Saito
et al.~\cite{saito}); DCO$^+$ (Williams et al.~\cite{williams98}).
However, the $70\hbox{--}90''$ beam size of most of these studies is a
severe limitation as the largest D/H enhancements are expected on
small scales, where densities are high.
In L1544, heavy depletion occurs in
a region 13000~AU across -- only $90''$ diameter in nearby Taurus (Caselli et
al.~\cite{caselli99}).  In order to prove a general link between
accretion and high D/H ratios, observations are needed of molecular
D/H ratios in a sample of protostars with high enough resolution to
probe the depletion regions.

I present the results of such a survey of [NH$_2$D]/[NH$_3$] ratios
towards 11 protostars in Perseus, Taurus and Orion, observing
NH$_2$D with the IRAM~30m at $28''$ resolution (maximally 11000~AU at
the distance of Orion) and centering the observations on the
protostellar dust peaks where the strongest depletion is likely to be
found.

\section{Observations}


NH$_2$D observations were made at the IRAM 30m on Pico Veleta in
May~2000 and June~2002 towards 11~protostellar objects in Perseus,
Taurus and Orion (Table~\ref{tbl:sourcelist}).  
I observed the 85.93~GHz 1(1,1)-1(0,1) ortho transitions (6 hyperfine
lines)(see. Olberg et al.~1985).  The 1(1,1) state lies 20.7~K above
ground.  Integration times were 10 minutes (position switched) and 20
minutes for the two Orion sources; system temperatures were 165--190~K
with typical summer conditions of 4mm~PWV (precipitable water vapour).
Observations of the two NGC~1333 sources in June~2002 were made under
7mm~PWV (T$_{\rm sys} =$~260--290~K).  Noise levels were 50--90~mK on
40~kHz channels.  Spectra were corrected for the beam efficiency
$B_{\rm eff} = 0.73$.
\begin{table}
   \caption[] {Source positions and velocities} 
\label{tbl:sourcelist}

   \begin{flushleft}
   \begin{tabular}{l r@{~~}c@{~~}c c@{~~}c@{~~}c r}
   \hline
   \noalign{\smallskip}
   Object   &  \multicolumn{3}{c}{RA (J2000) }
            &  \multicolumn{3}{c}{Dec (J2000) }
	    & $v_{\rm LSR}$  \\
   	    &  [ $^{\rm h}$  &  $^{\rm m}$  &  $^{\rm s}$\ ]
            &  [ $\degr$     &  $\arcmin$     &  $\arcsec$ ]
	    &  km~s$^{-1}$ \\
   \hline
   \noalign{\smallskip}
{\em Perseus}&&&&&&&\\
B5IRS1   &03 & 47 & 41.3   &$+$32 & 51 & 42  &10.2\\
L1448mms &03 & 25 & 38.8   &$+$30 & 44 & 05  & 4.7\\
L1448NW  &03 & 25 & 35.6   &$+$30 & 45 & 34  & 4.7\\
HH211    &03 & 43 & 56.8   &$+$32 & 00 & 50  & 9.2\\
IRAS03282$+$3035 &03 & 31 & 20.4   &$+$30 & 45 & 25  & 7.1\\
NGC1333~IRAS4A  &03 &26 & 04.8  &$+$31 &03 &14 &7.0\\
NGC1333~DCO$^+$ &03 &26 &06.6  &+31 &03 &08 &6.5\\
{\em Taurus}&&&&&&&\\
L1551IRS &04 & 31 & 34.1   &$+$18 & 08 & 05  & 6.2\\
L1527    &04 & 39 & 53.5   &$+$26 & 03 & 05  & 5.7\\
{\em Orion}&&&&&&&\\
RNO43    &05 & 32 & 19.4   &$+$12 & 49 & 32  &10.4\\
HH111    &05 & 51 & 46.3   &$+$02 & 48 & 30  & 8.5\\
   \noalign{\smallskip}
   \hline
   \end{tabular}
   \end{flushleft}
%
\end{table}

\begin{figure}
   \centering
   \includegraphics[width=180pt]{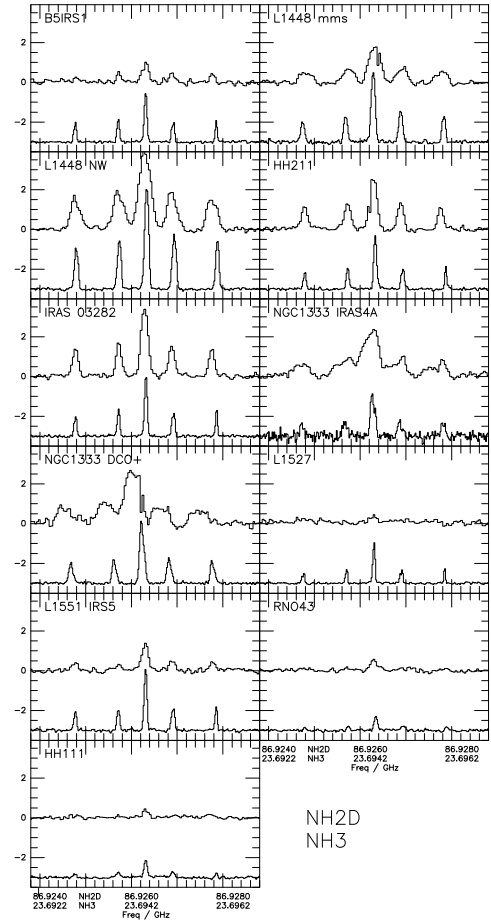}
	\caption{NH$_2$D (upper) and NH$_3$ (lower) spectra.  The
	   temperature scale is $T_{\rm MB}$ in Kelvin.  Note that the
	   NH$_2$D lines appear wider because of the factor 3.6
	   frequency difference between the NH$_3$ and NH$_2$D transitions.}
	   \label{fig:nh2dspec}
   \end{figure}

The main NH$_2$D line was detected in all 11 sources.  In the seven
Perseus sources, the four outlying hyperfine transitions were also
detected with peak brightness above 5$\sigma$.  In L1551IRS5, the
hyperfine transitions were above 3$\sigma$.  In the other Taurus and
Orion sources, only the main hyperfine line was detected.  Spectra are 
shown in Fig.~\ref{fig:nh2dspec}.


NH$_3$ observations at 24~GHz were made at the Effelsberg 100m in
May~2002 using the 8192-channel spectrometer, which allows observation
of the $(J,K)=(1,1)\hbox{--}(4,4)$ metastable transitions
simultaneously.  System temperatures were 40--50~K.  Calibration was
checked on W3(OH) and spectra were converted to Kelvin.
The Effelsberg beamsize at 24~GHz is $37''$.  NH$_3 (1,1)$ was detected in
all sources; the NH$_3 (1,1)$ hyperfines and the (2,2) main line were
detected in all sources except the two Orion sources (RNO43 and
HH111).  The upper state energies for the (1,1) and (2,2) lines are
23.8 and 65.0~K respectively.  The (3,3) and (4,4) lines were not
detected.

\section{Analysis and results}


%
%

NH$_2$D observations for the Perseus sources with well-detected
hyperfines were reduced using CLASS method HFS to fit the six
hyperfine lines (see Tin\'e et al.~(2000) for frequencies and line
strengths) and derive optical depths and linewidths
(Table~\ref{tbl:linepars}).  Where not all the hyperfines were
detected, or the HFS fit gave too great an uncertainty in the optical
depth, the main line was fitted with a single Gaussian (L1527,
L1551IRS5, RNO43, HH111).

NH$_3 (1,1)$ observations were reduced using CLASS method NH$_3$(1,1)
to fit the hyperfine structure and derive optical depths and
linewidths (Table~\ref{tbl:linepars}).  Excitation temperatures in the
(1,1) line were 5--10~K assuming unity filling factor, but as the
filling factor is uncertain these temperature estimates are of limited
value.  More usefully, I derived the rotation temperature $T_{12}$
using the method of Bachiller et al.~(\cite{bachiller87}), fitting the
(2,2) main line with a single Gaussian.  These temperatures are
independent of filling factor and calibration.  Rotation temperatures
were $T_{12} = 11-14$~K in the Taurus and Perseus sources.  At
these low temperatures, the rotational temperature is a good
approximation to the kinetic temperature $T_{12} \simeq T_{\rm kin}$
(Walmsley \& Ungerechts~\cite{walmsleyungerechts83}).  For RNO43 in
Orion, $T_{12}=14.6~K$.  The NH$_3(2,2)$ line in the other Orion source,
HH111, was too weak for an estimate of $T_{12}$.

For the Perseus sources, optical depths were available from hyperfine
fits, and column densities and [NH$_2$D]/[NH$_3$] ratios were derived
assuming LTE at $T_{\rm kin}=T_{12}$.  The resulting [NH$_2$D]/[NH$_3$] ratios are 0.17--0.33 (Table~\ref{tbl:ratios}).


For the Taurus sources L1527 and L1551~IRS5, optical depths were
available for NH$_3$ but not for NH$_2$D.  NH$_2$D beam-averaged
column densities were derived from the main hyperfine integrated
intensities at $T_{\rm kin}=T_{12}$.  The low [NH$_2$D]/[NH$_3$]
ratios, 0.04--0.05, likely reflect the fact that the NH$_2$D column
densities are beam-averaged whereas the NH$_3$ optical depths refer to
higher column density material on smaller scales, so the ratios are
lower limits.

\begin{table*}
\caption{NH$_2$D and NH$_3$ observed line parameters.  Uncertainties
(in brackets) come from either formal errors on optical depth from the hyperfine fits or 30\% absolute calibration uncertainties where column density was derived from integrated intensity.}
\label{tbl:linepars}
\begin{tabular}{l l l l l l l l}
\hline
Source	&\multicolumn{3}{c}{NH$_2$D 85.93~GHz}	
	&\multicolumn{3}{c}{NH$_3 (1,1)$} 
	&NH$_3 (2,2)$\\
	&$\tau$ &$\int T_{\rm MB}^*\,dv$ &$\Delta v$ 
	&$\tau$ &$\int T_{\rm MB}^*\,dv$ &$\Delta v$ 
	&$\int T_{\rm MB}^*\,dv$\\
	&		&K km s$^{-1}$			&km s$^{-1}$ 
	&		&K km s$^{-1}$			&km s$^{-1}$ 
	&K km s$^{-1}$	\\
\hline
B5IRS1	&3.22 (0.74)	&--		&0.46 (0.03)
	&3.74 (0.06)	&--		&0.58 (0.00)		
	&0.27 (0.02)\\
L1448mms&2.43 (0.29)	&--		&0.88 (0.02)
	&2.79 (0.04)	&--		&1.00 (0.00)	
	&1.23 (0.05)\\
L1448NW	&3.66 (0.12)	&--		&0.83 (0.01)
	&5.37 (0.04)	&--		&0.89 (0.00)
	&1.41 (0.03)\\
HH211	&4.54 (0.28)	&--		&0.49 (0.01)
	&2.70 (0.15)	&--		&0.68 (0.01)
	&0.53 (0.03)\\
IRAS03282&3.39 (0.18)	&--		&0.50 (0.01)
	&3.53 (0.15)	&--		&0.59 (0.01)
	&0.39 (0.02)\\
NGC1333~IRAS4A &1.46 (0.29)	&--	&1.38 (0.04)
	&1.66 (0.50)	&--		&1.28 (0.06)
	&0.78 (0.15)\\
NGC1333~DCO$^+$ &1.84 (0.25)	&--	&1.19 (0.02)
	&2.54 (0.08)	&--		&1.20 (0.01)
	&0.95 (0.04)\\
L1527	&--		&0.247 (0.074)	&0.83 (0.13)
	&1.51 (0.01) 	&--		&0.50 (0.01)
	&0.15 (0.02)\\
L1551IRS5&--		&0.680 (0.204)	&0.60 (0.03) 
	&2.14 (0.02)	&--		&0.66 (0.01)
	&0.52 (0.03)\\
RNO43	&--		&0.301 (0.090)		&0.68 (0.07) 
	&--		&0.791 (0.237)		&1.04 (0.04)
	&0.18 (0.02)\\
HH111	&--		&0.177 (0.053)		&0.51 (0.07) 
	&--		&0.964 (0.289)		&1.03 (0.04)
	&--\\
\hline
\end{tabular}
\end{table*}


For the Orion sources RNO43 and HH111, NH$3 (1,1)$ optical depths were
much less than one and very uncertain, so the NH$_3$ beam averaged
column density was calculated from the integrated intensity of the
main hyperfine line.  NH$_2$D and NH$_3$ column densities were
calculated at 14.6~K (the measured $T_{12}$ in RNO43).  The ratios are
additionally divided by the ratio of the beam areas to give
[NH$_2$D]/[NH$_3$] ratios of 11-22\%.  This assumes that the source is
centrally condensed and N(NH$_3$) is higher on the NH$_2$D beam
scales.  The correction is conservative, and the [NH$_2$D]/[NH$_3$]
ratios for the Orion sources could be a factor of 1.75 higher if the
NH$_2$D is extended, or 19--39\%.


When calculated without optical depth information the ratios are less
certain, as the absolute calibration uncertainty at each frequency is
$\sim 30\%$.  In contrast, where optical depths are measured the
[NH$_2$D]/[NH$_3$] ratio does not depend on the absolute calibration.
All the column density determinations are of course limited by the
lack of information about source structure from single spectra.
Foreground self-absorption and multiple components (which particularly
affect NGC1333) and the beam size difference between Effelsberg and
the 30m cannot be fully accounted for with the present data.

\begin{table}
\caption{Temperatures, column densities and [NH$_2$D]/[NH$_3$] ratio.}
\label{tbl:ratios}
\begin{tabular}{l l l l l}
\hline
Source	&$T_{12}$  
	&$N(\hbox{NH}_2\hbox{D})$	&$N(\hbox{NH}_3)$ &[NH$_2$D]/\\
 	&K
	&$\times 10^{14} \hbox{ cm}^{-2}$&$\times 10^{14} \hbox{ cm}^{-2}$  &[NH$_3$]\\
\hline
B5IRS1		&10.8	&2.12(0.49)		&13.4(0.2)	&0.18(0.04)\\
L1448mms	&13.7	&3.95(0.47)		&18.0(0.3)	&0.20(0.02)\\
L1448NW		&11.7	&4.74(0.16)		&29.7(0.2)	&0.17(0.01)\\
HH211		&12.8	&3.85(0.24)		&11.6(0.6)	&0.33(0.03)\\
IRAS03282	&11.3	&2.52(0.13)		&12.9(0.6)	&0.22(0.01)\\
NGC1333:		&&&&\\
IRAS4A		&14.2	&3.90(0.78)		&13.8(4.2)	&0.25(0.09)\\
DCO$^+$ 	&12.8	&3.76(0.51)		&19.2(0.6)	&0.19(0.03)\\
L1527		&10.8	&0.16(0.05)		&3.79(0.02)	&0.04(0.02)\\
L1551IRS5	&12.6	&0.42(0.12)		&8.88(0.07)	&0.05(0.01)\\
RNO43		&14.6	&0.18(0.05)		&0.46(0.14)	&0.22(0.10)\\
HH111		&--	&0.10(0.03)		&0.56(0.17)
&0.11(0.06)\\
\hline
\end{tabular}


\end{table}


\section{Discussion}
\label{sect:discussion}


In the Perseus sources where the ratios are best determined,
[NH$_2$D]/[NH$_3$]~$=$~17--33\%.  The highest ratios derived by Shah
\& Wootten~(\cite{sw01}) and Saito~et~al.~(\cite{saito}) are 13\% for
sources also on this sourcelist.  Differences in excitation assumptions
may also have an effect, but the lower ratios derived by Shah \&
Wootten~(\cite{sw01}) can easily be explained if the fractionation is
enhanced on small scales, towards the protostellar cores, as their
beam is larger (80--90$''$ compared to $27''$).  Saito et
al.~(\cite{saito})'s $18''$ resolution observation towards B1 is
$15''$ away from the dust peak (Matthews \& Wilson~\cite{matthews02})
and shows a lower ratio, which also supports the idea that
[NH$_2$D]/[NH$_3$] reduces away from the cores.  (No conclusion can be
drawn from their uncertain NH$_2$D column in L1448).  The ratios for
protostars are also higher than those towards L134N, presumably
because the protostars are denser and more depleted
([NH$_2$D]/[NH$_3$]$=0.1$; Tin\'e et al.~\cite{tine00}).

Why should [NH$_2$D]/[NH$_3$] be higher close to the protostars?  Is
cold gas-phase chemistry in the presence of freezeout enhancing
ratios in the cores, or are highly fractionated grain ices evaporating 
near the young stars?  


Chemical models indicate that [NH$_2$D]/[NH$_3$] ratios of 30\% can be
produced by ion-molecule chemistry if heavy molecules are depleted
(see Fig.~3 of Roberts \& Millar~\cite{rmdoubled}; Fig.~2 of Rodgers
\& Charnley~\cite{rcnd2h}).  The ratios can be even higher if
branching ratios for dissociative recombination favour deuterium
retention (Lis et al.~2002).  The corresponding fractionation required
in molecular ions is similar or less than that already measured in
L134N (Rodgers \& Charnley~\cite{rcnd2h}; N$_2$D$^{+}$ observations by
Tin\'e et al.~\cite{tine00}).
Gas-phase chemistry with freezeout and depletion
can therefore relatively easily explain the high [NH$_2$D]/[NH$_3$]
ratios.


An alternative hypothesis is that removal of highly fractionated grain
mantles enhances ratios.  Infrared observations suggest that
NH$_3$ may be present in ices, though the level is still uncertain
(Lacy et al.~\cite{lacy98}; Gibb et al.~\cite{gibb00}; Dartois \&
D'Hendecourt~\cite{dartois01}), but NH$_3$ ice alone evaporates at
60--70~K; in a water matrix temperatures greater than 90~K are
required.  The temperatures of our sources, which are mostly Class~0
protostars, rule out much thermal ice evaporation.  Only B5IRS1 and
L1551 have bolometric (dust) temperatures above 60~K.  The source with
the highest [NH$_2$D]/[NH$_3$] ratio, HH211, is probably extremely
young (Gueth \& Guilloteau~\cite{guethguilloteau99}) and has a
bolometric temperature of only 30~K.  Kinetic temperature estimates
for the NH$_3$ are all less than 15~K.

The possibility remains that highly fractionated grain mantles are
being removed by non-thermal evaporation, ie. slow, non-dissociative
shocks.  [NH$_2$D]/[NH$_3$] ratios are around 10\% in regions where
grain evaporation is known to be occuring (hot cores and IRAS~~16293;
van Dishoeck et al.~\cite{vandishoeck95}),
and may originate in the ices, as may high [D$_2$CO]/[H$_2$CO] ratios
(Ceccarelli et al.~1998,2002).  Enhancements of
another molecule, CH$_3$OH, in a similar sample of cores, are also
attributed to non-thermal ice evaporation (Buckle \&
Fuller~\cite{bucklefuller02}).  Two conditions must be satisfied if a
non-thermal removal mechanism operates.  Firstly, it must reproduce
the low linewidths of the NH$_3$ ($<1$~km~s$^{-1}$) and NH$_2$D
(linewidths the same or lower), which suggest that NH$_2$D is tracing
quiescent rather than shocked material.  Secondly, if the release of
ice material is to explain all the NH$_2$D observed, the
[NH$_2$D]/[NH$_3$] ratio in the ices must be as high or higher than
that measured in the gas phase, because of dilution by low
[NH$_2$D]/[NH$_3$] ratios in regions of the core without ice
evaporation.


Observations suggest that, in these sources, ices can only be released
in a small fraction of the core, otherwise CH$_3$OH enhancements of
several hundred percent would be expected, as seen in outflows and hot
cores, rather than the moderate 10\% observed (Buckle \&
Fuller~\cite{bucklefuller02}), and column densities of NH$_3$ would be
enhanced by orders of magnitude over the surrounding molecular clouds,
which is not the case (Bachiller et al.~\cite{bachiller87},
\cite{bachiller91}, \cite{bachiller93}; Benson \&
Myers~\cite{bensonmyers89}; Anglada et al.~\cite{anglada89}).  


To obtain the observed ratios even without dilution, the statistical
treatment of Rodgers \& Charnley~(\cite{rcnd2h})
following Brown \& Millar~(\cite{brownmillar89a};\cite{brownmillar89b})
requires an atomic D/H ratio in the gas phase at the time of freezeout
of 0.8--1.6.  This is at the limit of what either steady-state or depletion
models predict, as these cannot produce atomic D/H~$> 0.1$ (Roberts \&
Millar~\cite{rmsingled}; Roberts, priv. comm.).


With dilution, therefore, current theories of grain mantle chemistry
cannot explain the NH$_2$D enhancement, but would require
either higher [D]/[H] ratios during freezeout or preferential
formation of deuterated molecules in the ice.


The [NH$_2$D]/[NH$_3$] ratios of $\sim 0.2$ in NGC1333 is consistent with
ND$_3$ column densities 
(van der Tak et al.~\cite{vdt02}; Lis
et al.~\cite{lis02}), using the models of Rodgers \&
Charnley~(\cite{rcnd2h}).  This is in contrast to earlier
[NH$_2$D]/[NH$_3$] ratios of $10\%$ which predicted [ND$_3$]/[NH$_3$]
an order of magnitude lower than measured.  
%
As the theory predicts
[ND$_3$]/[NH$_3$]$\propto$[NH$_2$D]/[NH$_3$]$^3$, ND$_3$ predictions
are very sensitive to [NH$_2$D]/[NH$_3$] ratios, and more observations
of both isotopomers are needed to say if the theories for multiple
deuteration are correct.  Measurements of ND$_2$H in the same sources
would provide a further test: [ND$_2$H]/[NH$_3] \sim 0.03$ is expected
from the models, for the given [NH$_2$D]/[NH$_3$] ratios.  Detailed
maps of individual sources would also test the relationship between
high deuteration and CO depletion.

\section{Summary and conclusion}

[NH$_2$D]/[NH$_3$] ratios in low mass protostellar cores on 10,000~AU
scales can reach 30\% percent.  Such high ratios can be simply
explained by recent models of cold gas-phase chemistry with depletion
due to freezeout (Roberts \& Millar~\cite{rmsingled},
\cite{rmdoubled}).  Current models of ice formation even under
conditions of high depletion have difficulty explaining the high
fractionation observed.

\begin{acknowledgements}
      The author would like to thank Malcolm Walmsley, Floris van der
      Tak and Helen Roberts for useful discussions.  This work was
      supported by the Deut\-sche
      For\-schungs\-ge\-mein\-schaft SFB 494.
\end{acknowledgements}

\end{document}